\documentstyle[aasms,epsfig,twoside]{article}

\newlength{\mpar}
\begin{document}

%%%%%%%%%%%%%%%%%%%%%%%% TITLE PAGE  %%%%%%%%%%%%%%%%%%%%%%%%%%%%%%%%
%\twocolumn[$\;$\vspace{0.5cm}

\title{Upper limit to $\Omega_B$ in scalar-tensor gravity theories}

\author{\sc Jean-Michel Alimi \& Arturo Serna} 
\affil{Laboratoire d'Astrophysique Extragalactique et de Cosmologie,
CNRS URA 173, Observatoire de Paris-Meudon, 92195 Meudon, France}

\vspace{0.1cm} 
%%%%%%%%%%%%%%%%%%%% ABSTRACT & SUBJECT HEADINGS %%%%%%%%%%%%%%%%%%%%%
%\begin{center}\begin{minipage}{15cm}

\begin{abstract}
In a previous paper (Serna \& Alimi 1996b), we have pointed out the 
existence of some particular scalar-tensor gravity theories able to  relax
the nucleosynthesis constraint on the cosmic baryonic density.  In this
paper, we present an exhaustive study of primordial  nucleosynthesis in the
framework of such theories taking into  account the currently adopted
observational constraints. We show  that a wide class of them allows for a
baryonic density very close to  that needed for the universe closure. This
class of theories  converges soon enough towards General Relativity and,
hence, is  compatible with all solar-system and binary pulsar gravitational 
tests. In other words, we show that primordial nucleosynthesis does  not
always impose a very stringent bound on the baryon contribution  to the
density parameter. \vspace{0.1cm}
\end{abstract}

\pagestyle{myheadings} \markboth{\centerline{\uppercase{Alimi and
      Serna.}}} {\centerline{\uppercase{Upper limit to $\Omega_B$ in
      scalar-tensor gravity theories}}}

\thispagestyle{empty} 

%\end{minipage}\end{center}
\vspace{0.5cm} %]
%%%%%%%%%%%%%%%%%%%%%%%% TEXT & ACKNOWLEDGMENTS %%%%%%%%%%%%%%%%%%%%%%

\section{INTRODUCTION}

One of the most convincing pieces supporting the standard hot big bang
model for the early Universe is the excellent agreement between the
predicted and observed primordial light element abundances. However,
this agreement requires that the present baryon density $\rho_b$ must
be smaller (Walker et al. 1991, Copi, Schramm \& Turner 1995a-1995b,
Olive 1996), than about 10\% of that needed for the Universe closure
($\Omega_B\equiv\rho_b/\rho_{crit} < 0.1 $). A high-density universe
($\Omega \gg 0.1$), seems then to be impossible unless that most dark
matter in the Universe were non-baryonic.

The problem in admitting large $\rho_b$ values is that, in the standard
primordial nucleosynthesis model, they lead to the overproduction of $^4$He
and $^7$Li and the underproduction of deuterium. In order to understand this
difficulty, we recall that primordial nucleosynthesis starts, in the early
Universe, soon after that the cosmic temperature becomes smaller than that
needed to maintain the proton-to-neutron ratio in its equilibrium value
(freezing-out temperature). The nuclear reactions which then take place lead
first to the D and $^3$He formation. These elements are then burnt to
produce $^4$He. At the end of this process, essentially all neutrons are
incorporated into $^4$He, while just a small fraction of them remains in D
and $^3$He. Consequently, the $^4$He abundance mainly depends on the
fraction of neutrons existing at the beginning of the nucleosynthesis
process or, equivalently, on the freezing-out temperature for the $n/p$
ratio. On the other hand, since the burning of D and $^3$He grows with the
baryon density, the final yields of these two elements decrease with $\rho_b$%
, while those of $^4$He and $^7$Li increase%
\footnote{Although the  lithium-7 history is  much more
  complicated, when the baryon density  is relatively large, its final
  abundance also increases with $\rho_b$  due to the $^7$Li production
  through $^7$Be}.

Several attempts to relax the nucleosynthesis bound on $\Omega_B$ have been
previously considered (Malaney \& Mathews 1993). This is the case, for
example, of inhomogeneous big bang models, and the attempts of modifying the
nuclear reaction rates by the introduction of new decaying particles (e.g.
Schramm 1991, Gyuk \& Turner 1994). However, all these attempts have been
unsuccessful or seem to be mere modifications to try to solve this problem.
The nucleosynthesis bound on $\Omega_B$ is today considered as a
unavoidable constraint which must be imposed even to study the formation of
large-scale structures in the Universe.

Another possibility also analyzed in the literature consists of modifying
the gravity description and, hence, the Universe expansion rate without
altering the nuclear physics, the Universe composition or the conservation
laws. This last possibility can be considered as more natural from a
theoretical point of view because it would allow us to construct a global
framework which is coherent with that often assumed to describe the epochs
prior to primordial nucleosynthesis. As a matter of fact, to study the
earliest times in the Universe evolution, the existence of an extra scalar
field, in addition to the Einstein metric tensor, is usually assumed. Such
gravity theories are termed scalar-tensor theories (Bergmann 1968, Wagoner
1970, Nordtvedt 1970), and are the simplest generalization of the Einstein
theory. They provide a natural (non-fine-tuned) way to restore the original
ideas of inflation while avoiding the cosmological difficulties coming from
the vacuum-dominated exponential expansion obtained in General Relativity
(GR) (La \& Steinhardt 1989, Weinberg 1989, Steinhardt \& Accettta 1990,
Barrow \& Maeda 1990, Liddle \& Wands 1992, Deruelle, Gundlach \& Langlois
1992, Garcia-Bellido \& Wands 1992, Barrow 1995). Scalar-tensor theories
also arise in the current theoretical attempts at deepening the connection
between gravitation and the other interactions. For example, in modern
revivals of the Kaluza-Klein theory and in supersymmetric theories with
extra dimensions, one or more scalar fields arise in the compactification of
these extra dimensions (Jordan 1949, Kolb, Perry \& Walker 1986, Cho 1992,
Wesson \& Ponce de Leon 1995). Furthermore, scalar-tensor theories may also
appear as a low-energy limit of superstring theories (Green, Schwarz \&
Witten 1988).

The Big Bang nucleosynthesis (BBN) in presence of a scalar field has been
previously analyzed in some cases (Bekenstein \& Meiseils 1980, Arai,
Hashimoto \& Fukui 1987, Serna, Dominguez-Tenreiro \& Yepes 1992, Serna,
Dominguez-Tenreiro 1993, Serna \& Alimi 1996b). Among all these analyses,
only some models pointed out by Serna \& Alimi (1996b) were able to relax
the nucleosynthesis bound on $\Omega_B$. The key of such a result was that
the expansion rate resulting in such theories avoided simultaneously the
overproduction of $^4$He and $^7$Li and the underproduction of D. In this
paper we construct a wide class of scalar-tensor gravity theories also
allowing a large range for baryonic density.

\section{SCALAR-TENSOR NUCLEOSYNTHESIS}

Scalar-tensor theories are characterized by an arbitrary coupling function 
$\omega(\Phi)$, which determines the relative importance of the additional
scalar field $\Phi$. GR is the asymptotic case in which the coupling
function is infinite and $\Phi =$ constant = 1. Since GR reproduces very
accurately the solar-system and binary pulsar dynamics (Will 1993), the
behavior of any viable scalar-tensor theory must be restricted to be at
present extremely close to that implied by GR. The physical conditions in
the early Universe and the subsequent cosmological evolution in the
framework of these theories can be nevertheless very different from the
usual ones.

In order to analyze the possible existence of scalar-tensor theories,
able to reproduce the right primordial abundances for an $\Omega_B$
interval much wider than in GR, we must first calculate the resulting
cosmological models.  To that end, we consider an {\bf isotropic and
  homogeneous} universe and a specific form for the coupling function.
The line element has then a Robertson-Walker form and the
energy-momentum tensor corresponds to that of a perfect fluid. In that
concerning the coupling function, a convenient form is given by
\begin{equation}
\label{eq:w32}\mid\!3+2\omega\!\mid = (3/\lambda^{2})(x^{-1}+k)  
\end{equation}
where $\lambda$ and $k$ are arbitrary constants and $x =
\mid\!\Phi-1\!\mid$.  As a matter of fact, as shown by Serna \& Alimi
(1996a), such a form gives an exact representation for most of the
particular scalar-tensor theories proposed in the literature and, in
addition, it contains all the possible early behaviors of any theory
where $\omega(\Phi)$ is a monotonic, but arbitrary, function of
$\Phi$.

When such an $\omega(\Phi)$ function is introduced into the field equations,
one essentially obtains four different classes of scalar-tensor theories.
The two first classes correspond to singular models with a monotonic time
evolution of the speed-up factor ($\xi\equiv H/H^{FRW}$, where $H$ is the
Hubble parameter, while $H^{FRW}$ is that predicted by GR at the same
temperature). In the first class, the Universe expansion is always faster
than in GR ($\xi>1$), while the second class is characterized by an
expansion rate slower than in GR ($\xi<1$). Nonsingular models or models
with a critical temperature where $3+2\omega=0$ constitute the third class
of theories. Finally, the fourth class corresponds to models with a
non-monotonic $\xi(T)$ function. These last theories have an initial phase
where the Universe expansion is slower than in GR but, afterwards, it
becomes faster than in the standard cosmology. Obviously, the Universe
evolution in the framework of any viable scalar-tensor theory must satisfy 
$\xi\rightarrow 1$ at present.

In the first three classes of scalar-tensor theories, the right primordial
yields are only obtained (Serna \& Alimi 1996b) in the limit where such
theories are almost indistinguishable from GR, from the beginning of
primordial nucleosynthesis up to the present. Consequently, the allowed 
$\Omega_B$ interval is essentially the same as that implied by GR. 

We will then focus on the primordial production of light elements in
the framework of theories with a non-monotonic evolution of the speed-up
factor (class-4 models). This class of models is inevitably obtained when 
$\lambda /\sqrt{k}>1$, and both $3+2\omega$ and $(\Phi -1)$ have positive
values. We show below that new constraints on the baryonic density parameter
can then be obtained. 

The computation of the primordial production of light
elements\footnote{Our code solves the scalar-tensor cosmological
  equations by using a $6^{th}$ order Runge-Kutta integration, while
  the primordial nucleosynthesis rate equations are integrated by the
  Beaudet \& Yahil (1977) scheme.  The outputs of this code were
  extensively tested and compared to those obtained by other authors,
  as well as to the analytical solutions known for some particular
  cases. It was then used in previous works as, e.g., Serna {\it et
    al.} (1992), Serna \& Dom\`{\i}nguez-Tenreiro (1992, 1993), Serna
  \& Alimi (1996a, 1996b)} has been performed by using the updated
reaction rates of Caughlan and Fowler (1988) and Smith {\it et al}
(1993). We have considered $N_\nu =3,\;T_0=2.73$K, $\tau _n=889s$,
$H_0=50$km s$^{-1}$Mpc$^{-1}$. Other runs have been also performed,
with $\tau _n=889.8\pm 3.6$s and $H_0=100$km s$^{-1}$Mpc$^{-1}$, in
order to test the robustness of our conclusions.  For each theory
defined by the parameters $\lambda$ and $k$, we have searched the
largest $\Omega_B$ values for which there exists a present value of
the coupling function $\omega_0$ (i) compatible with all solar system
experiments and (ii) leading to the observed primordial abundances
(Copi, Schramm \& Turner 1995b):
\begin{displaymath}
\begin{array}{c}
0.221\leq \mbox{Y}_P\leq 0.243 \\ \mbox{D/H}\geq 1.5\cdot 10^{-5} \\ 
(\mbox{D}+^3\mbox{He)/H}\leq 1.1\cdot 10^{-4} \\ 
0.7\cdot 10^{-10}\leq ^7\mbox{Li/H }\leq 3.5\cdot 10^{-10}
\end{array}
\end{displaymath}
Some of these largest upper bounds on $\Omega_B$ are shown as
isocontour lines on figure 1. The $\omega_0$ values satisfying the
two previous criteria are different for different theories and, in
particular, they are smaller when the upper bound on $\Omega_B$ are
larger (see table). The baryon contribution to the total density
parameter cannot be therefore arbitrarily high because, for too large
$\Omega_B$ values, the resulting cosmological models will not be then
compatible with solar system experiments, which require
$\omega_0>500$. The allowed interval found for $\Omega_B$ is
\begin{displaymath}
0.01\leq\Omega_B\leq 0.78
\end{displaymath}
When a wider observational range for the $^7$Li/H
abundance is considered ($^7$Li/H $\leq 6\cdot10^{-10}$, which
corresponds to a depletion by a factor of 4), this interval becomes  
\begin{displaymath}
0.01\leq \Omega_B\leq 0.98
\end{displaymath}

In order to illustrate more precisely our procedure, we plot on
figures 2, the light elements primordial abundances for GR and for two
particular theories appearing on figure 1 . These two theories are
defined by $\lambda^2=0.4$, log$_{10}(\lambda^2/k)=7.8$ and
$\omega_0=1.3\cdot 10^9$ (square symbol), $\lambda^2=0.2$,
log$_{10}(\lambda^2/k)=8.$ and $\omega_0=5.75\cdot 10^3$ (circle
symbol). On these figures we see clearly that a range of large
$\Omega_B$ values exists, where predicted primordial abundances are
all compatible with observations. Any smaller upper bound on
$\Omega_B$ than those seen on these figures, can be however obtained by
considering larger $\omega_0$ values. We note also that, in fact, the
most constraining light elements are $^4$He and $^7$Li. Any
modification on the observational limits on these two elements would
imply a different allowed range for $\Omega_B$. In particular, we
see from figure 2d, that an almost flat Universe ($\Omega_B \simeq 1$)
is permitted when the upper observational limit for $^7$Li/H is
$6\cdot 10^{-10}$.

\section{Discussion and Conclusions}
A homogeneous and isotropic universe with a baryonic density very
close to that needed for closure by baryons is
then possible in the framework of class-4 scalar-tensor theories even
when the Universe is composed solely by known particles.

Since $\omega_0$ is much higher than the minimum value needed to
assure compatibility with all post-Newtonian experiments, these
theories are at present very close to GR. However, their behavior at
early times is very different from the usual one and can imply a
universe expansion rate during the nucleosynthesis process several
hundred of times faster than in the FRW cosmology.

The achievement of such $\rho_{b}$ values can be explained in the
following way. The Universe expansion rate at the beginning of
primordial nucleosynthesis is slower than in GR. This implies a
smaller freezing-out temperature and, hence, a tendency in these
theories to underproduce $^4$He.  The smaller $\omega_0$, the larger
this trend is. The $^4$He underproduction can be balanced by
considering larger $\rho_{b}$ values which in principle could imply an
excessive D burning. But, contrary to all other attempts of modifying
the standard hot big bang model, the Universe expansion rate obtained
in class-4 theories becomes during BBN faster than in GR.
Consequently, the D burning is not very effective because it occurs in
a shortest time and, hence, large $\rho_{b}$values are possible.

Obviously, in order to evaluate {\bf quantitatively} the light element
production in these theories, it was necessary to consider a specific
form for $\omega(\Phi)$(Eq. 1). However, this form already reveals a
large number of models for which $\Omega_B$ is much larger than in GR.
Furthermore, it is almost evident that such a result should be also
obtained for any other scalar-tensor theory, defined by a different
coupling function, but implying a similar non-monotonic behavior for
the speed-up factor. As a matter of fact, in order to avoid both an
overproduction of $^4$He and an underproduction of D, the only
condition is that the speed-up factor be smaller than unity at the
beginning of BBN and becomes, during the nucleosynthesis process,
larger than in GR.

It is important to note, that we do not claim here that the $\Omega_B$
value is necessarily high. The essential point that we want to stress
is that, in a homogeneous and isotropic universe composed only by
known particles, primordial nucleosynthesis does not always impose a
very stringent bound on the baryon contribution to the density
parameter.

Finally, it must be also pointed out that the study of inflation models in
the framework of these theories, where a form of the coupling function is
now known (Eq. 1), and their consequences on the primordial density
fluctuation spectrum, as well as on the formation of large-scale structures
are now important open questions which could lead to a new cosmological
scenario.

\newpage
\begin{table}
\begin{center}
{\small 
\begin{tabular}{rrrrrrrrrrrr}\hline\hline
\multicolumn{1}{c}{ } & \multicolumn{2}{c}{$\Omega_{B}^{max}=0.3$} &
\multicolumn{2}{c}{$\Omega_{B}^{max}=0.4$} &
\multicolumn{2}{c}{$\Omega_{B}^{max}=0.5$} &
\multicolumn{2}{c}{$\Omega_{B}^{max}=0.6$} &
\multicolumn{2}{c}{$\Omega_{B}^{max}=0.7$} &
\multicolumn{1}{c}{$\omega_{0}^{max}=500$} \\
\multicolumn{1}{c}{$\lambda^2$}  
&\multicolumn{1}{c}{$\kappa$}&\multicolumn{1}{c}{$\omega_{0}^{max}$}
&\multicolumn{1}{c}{$\kappa$}&\multicolumn{1}{c}{$\omega_{0}^{max}$}
&\multicolumn{1}{c}{$\kappa$}&\multicolumn{1}{c}{$\omega_{0}^{max}$}
&\multicolumn{1}{c}{$\kappa$}&\multicolumn{1}{c}{$\omega_{0}^{max}$}
&\multicolumn{1}{c}{$\kappa$}&\multicolumn{1}{c}{$\omega_{0}^{max}$}
&\multicolumn{1}{c}{$\kappa$} \\ \hline
0,20 & 0,98 & 1,46e12 & 1,45 & 8,80e12 & 2,11 & 1,03e13 & 4,20 & 2,02e11 & 6,17 & 4,00e08 & 8,714 \\
0,25 & 1,00 & 1,50e13 & 1,56 & 6,33e13 & 3,00 & 1,38e13 & 6,30 & 3,45e09 & 7,30 & 4,00e07 & 9,262\\
0,30 & 1,02 & 8,00e13 & 1,80 & 2,30e14 & 5,90 & 5,20e10 & 7,30 & 5,00e08 & 8,00 & 1,30e07 & 9,674\\
0,35 & 1,06 & 3,20e14 & 2,30 & 3,80e14 & 7,10 & 5,00e09 & 7,90 & 2,10e08 & 8,50 & 6,50e06 & 9,997\\
0,38 & 1,10 & 6,00e14 & 3,00 & 1,67e14 & 7,50 & 2,30e09 & 8,20 & 1,10e08 & 8,70 & 5,00e06 & 10,135\\
0,40 & 1,14 & 1,00e15 & 5,10 & 2,50e13 & 7,80 & 1,30e09 & 8,40 & 8,50e07 & 8,90 & 3,50e06 & 10,260\\
0,45 & 1,20 & 2,65e15 & 6,90 & 5,00e10 & 8,20 & 8,50e08 & 8,80 & 4,00e07 & 9,20 & 2,70e06 & 10,478\\
0,50 & 1,40 & 5,00e15 & 7,60 & 1,00e10 & 8,60 & 3,50e08 & 9,10 & 2,50e07 & 9,50 & 1,50e06 & 10,666\\
0,60 & 1,80 & 1,40e16 & 8,40 & 4,00e09 & 9,10 & 2,50e08 & 9,50 & 2,20e07 & 9,90 & 1,05e06 & 10,967\\
0,70 & 3,10 & 3,83e15 & 8,90 & 2,00e09 & 9,50 & 1,50e08 & 9,80 & 1,90e07 & 10,2 & 9,00e05 & 11,204\\
0,80 & 7,10 & 8,00e11 & 9,20 & 1,10e09 & 9,80 & 8,00e07 & 10,1 & 1,25e07 & 10,5 & 7,25e05 & 11,395\\
1,00 & 8,50 & 5,00e10 & 9,70 & 1,00e09 & 10,2 & 7,00e07 & 10,5 & 9,50e06 & 10,8 & 6,50e05 & 11,688\\ \hline
\multicolumn{12}{l}{{\footnotesize $\kappa\equiv\log_{10}(\lambda^2/k)$}}
\end{tabular}
}
\end{center}
\end{table}
%%%%%%%%%%%%%%%%%%%%%%%%%%%%%%%  REFERENCES  %%%%%%%%%%%%%%%%%%%%%%%%%%

\newpage
\begin{figure}
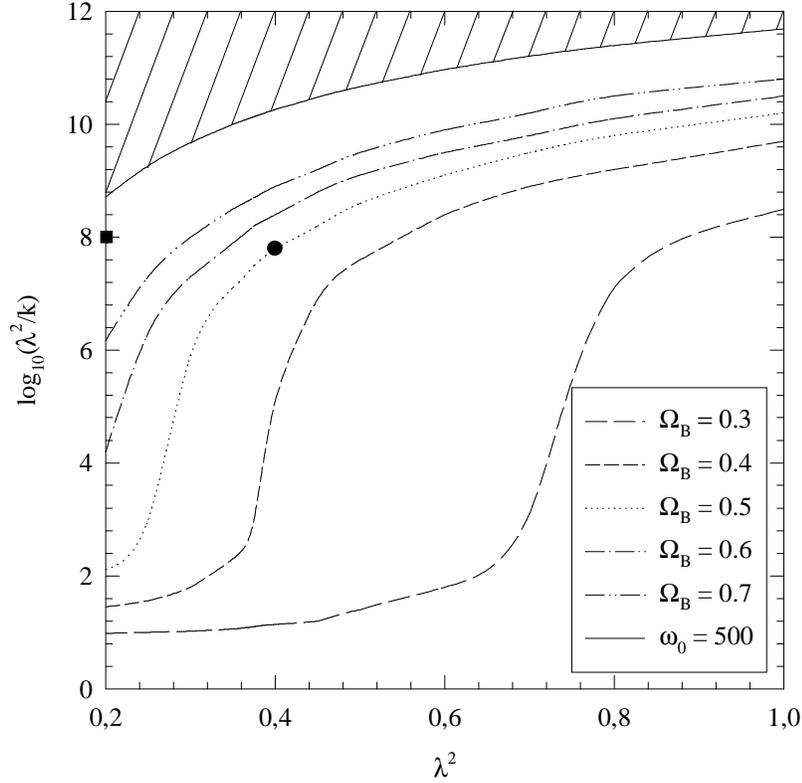

\caption{Space of theories leading to different upper bounds on $\Omega_B$: 
  The observational constraints used to construct this figure are:
  0.221$\leq$ Y$_p\leq$0.243, D/H$\geq1.5\cdot 10^{-5}$,
  (D+$^3$He)$\leq 1.1\cdot 10^{-4}$, $0.7\cdot 10^{-10}\leq$
  $^7$Li/H$\leq 3.5\cdot 10^{-10}$, and $h_0\in[0.4,1]$. The two
  symbols (square, circle) correspond to the two particular theories
  commented in the text and presented in figures 2. The dashed region
  represents the space of theories with an uppest limit on $\Omega_B$
  requiring an $\omega_0$ value smaller than 500.} \protect\label{omega}
\end{figure}
\begin{figure}
\caption{Primordial abundances of a) $^4$He, b) D/H, c) (D+$^{3}$He)/H, and
  d) $^7$Li, as a function of $\Omega_B$. The theories shown in this
  figure are defined by $\lambda^2=0.4$, log$_{10}(\lambda^2/k)=7.8$,
  $\omega_0=1.3\cdot 10^{9}$ (dashed line, corresponding to the circle
  symbol on figure 1), $\lambda^2=0.2$, log$_{10}(\lambda^2/k)=8$,
  $\omega_0=15.75\cdot 10^3$ (dotted line, the square symbol on figure
  1) and GR (solid line). A wider observational range for
  the $^7$Li/H abundance ($^7$Li/H $\leq 6\cdot10^{-10}$,
  dashed-dotted line) is also displayed}\protect\label{BBN}
\end{figure}

\newpage
\centerline{\epsfig{figure=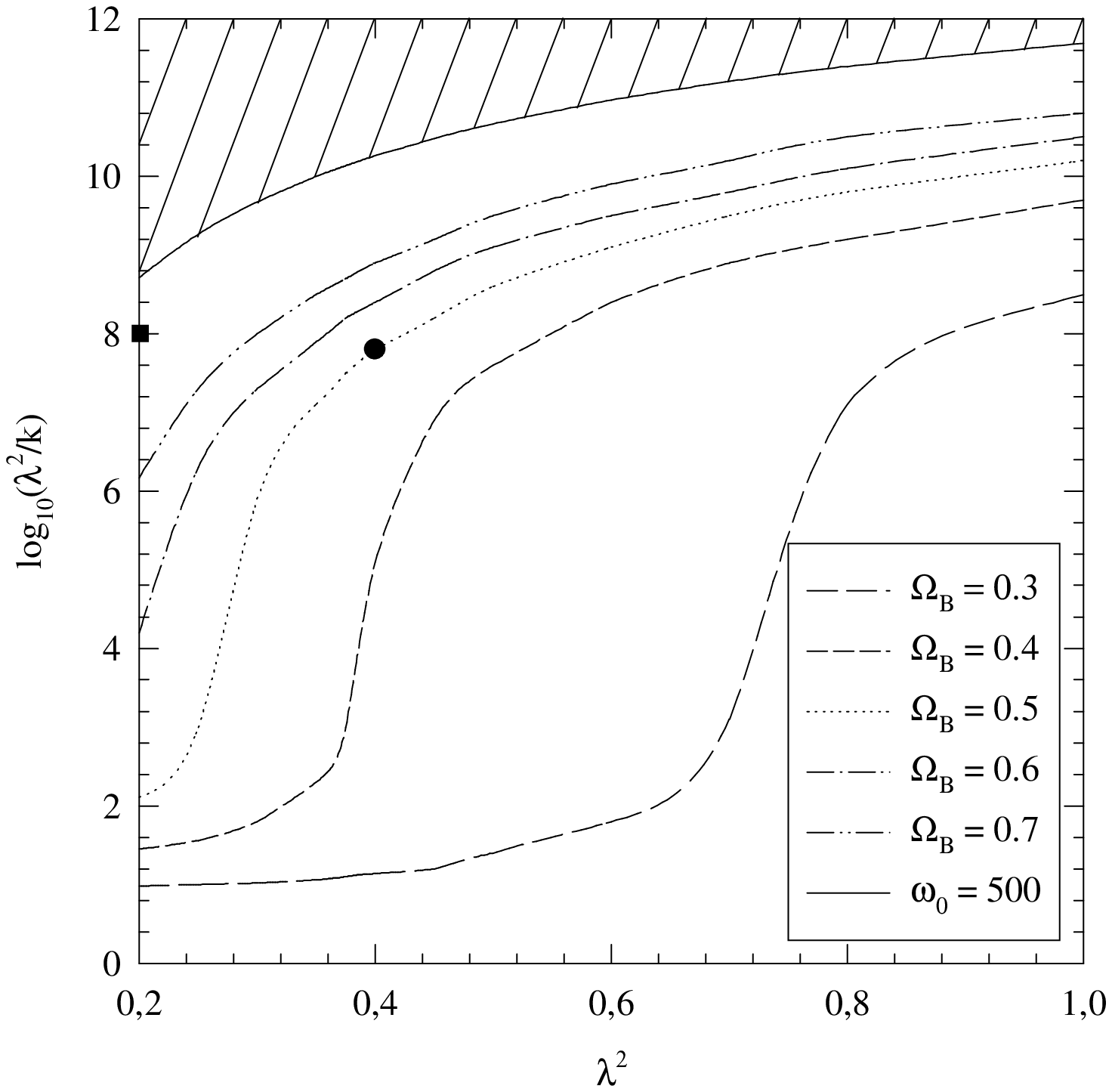,width=14cm}}
\newpage
\centerline{\epsfig{figure=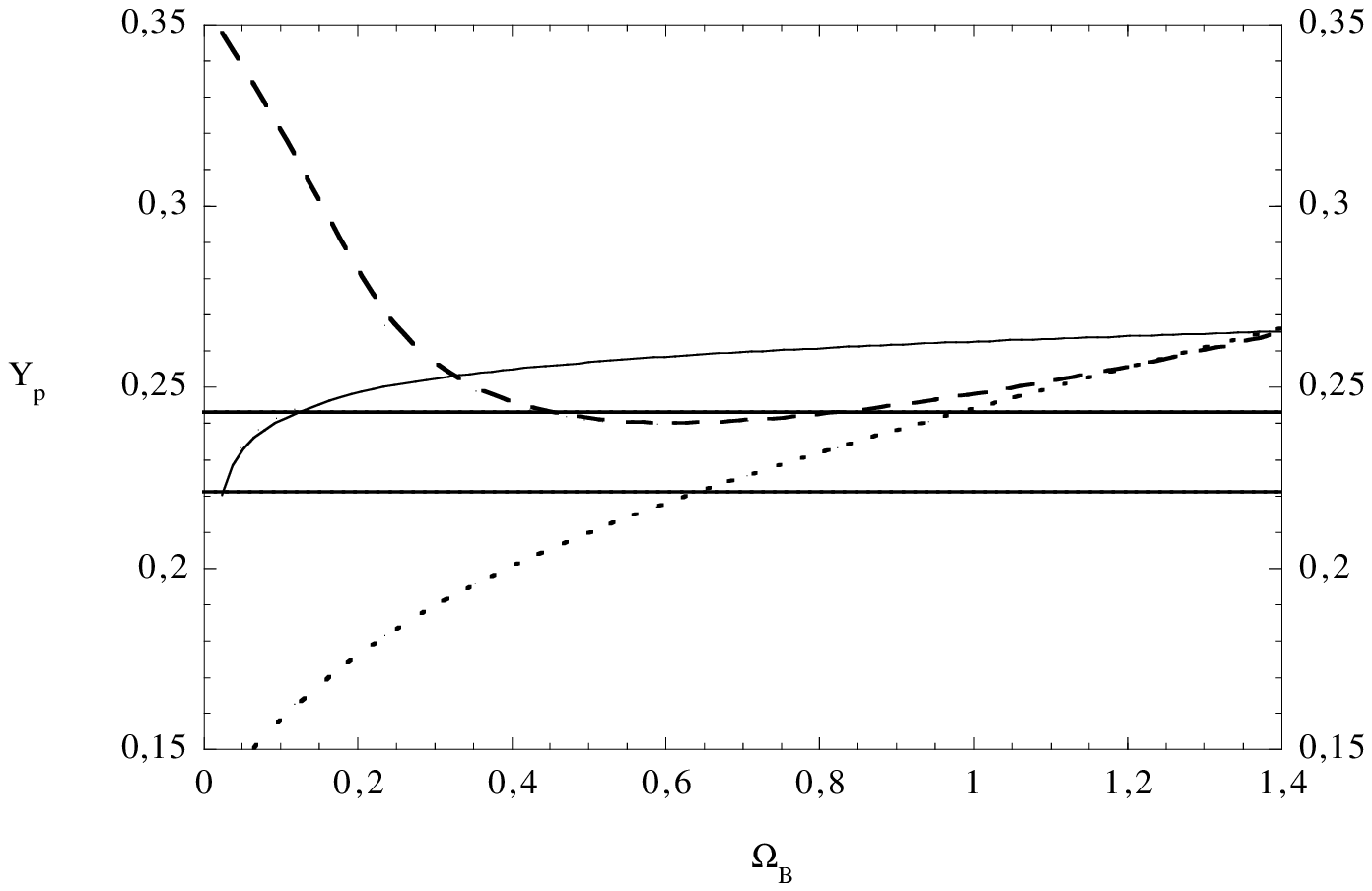,width=14cm}}
\newpage
\centerline{\epsfig{figure=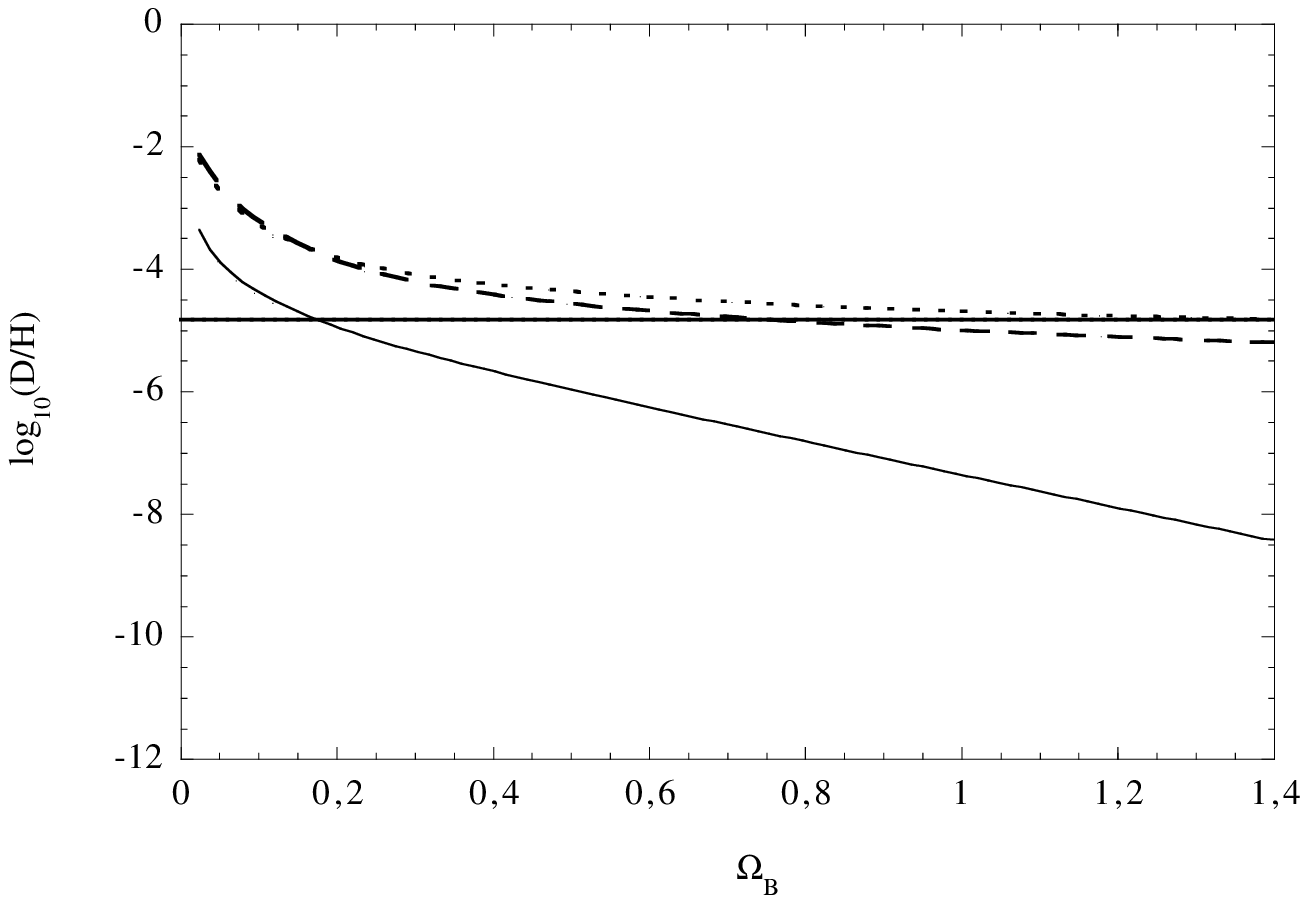,width=14cm}}
\newpage
\centerline{\epsfig{figure=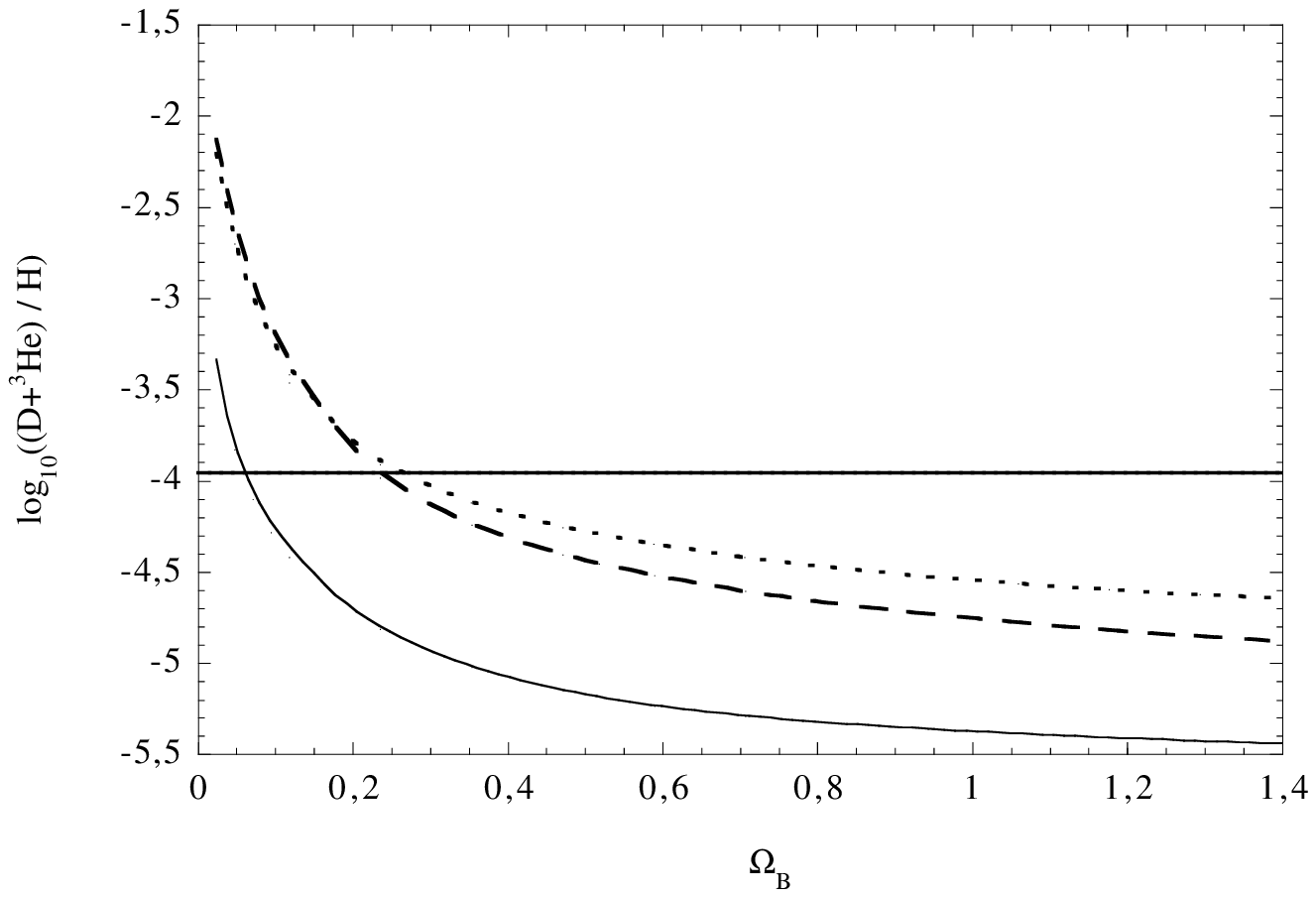,width=14cm}}
\newpage
\centerline{\epsfig{figure=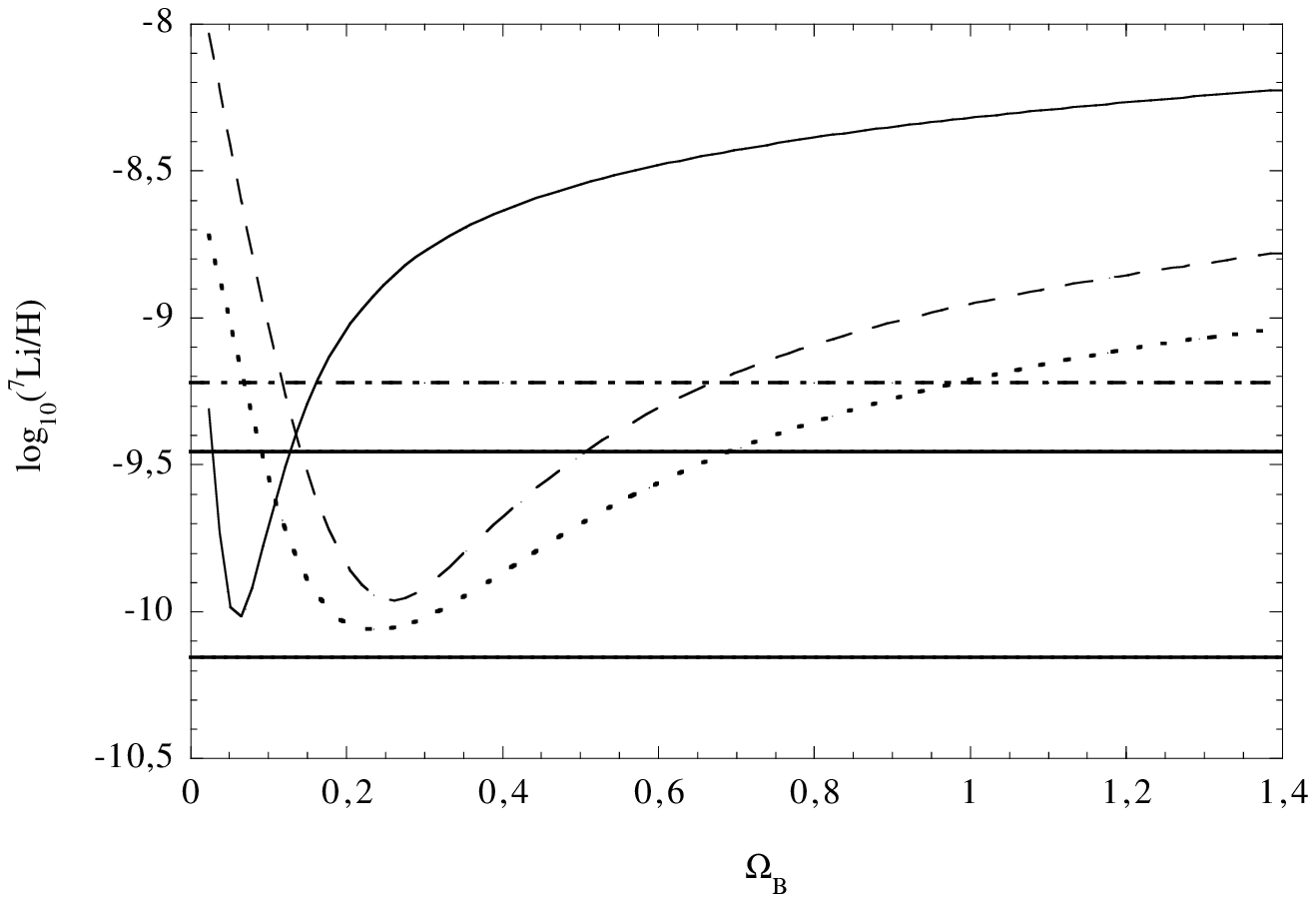,width=14cm}}
\end{document}